\documentclass[aps,showpacs,nofootinbib,superscriptaddress,twocolumn]{revtex4}
\usepackage{graphicx,subfigure}
\usepackage{amssymb}
\usepackage{bm}
\usepackage{array}
\usepackage{slashed}
\usepackage{braket}
\usepackage{multirow}
\usepackage{bigstrut}
\usepackage[nooneline,footnotesize,center]{caption2}
\usepackage{longtable}
\usepackage{hyperref}
\usepackage{amsmath}

\renewcommand{\d}{{\mathrm{d}}}

\begin{document}

\title{Nuclear EMC effect through $\bar{\Lambda}/\Lambda$ production in semi-inclusive deep-inelastic scattering processes}

\author{Chang Gong}
\affiliation{School of Physics and State Key Laboratory of Nuclear Physics and
Technology, Peking University, Beijing 100871, China}

\author{Bo-Qiang Ma}
\email{mabq@pku.edu.cn}
\affiliation{School of Physics and State Key Laboratory of Nuclear Physics and
Technology, Peking University, Beijing 100871, China}
\affiliation{Collaborative Innovation Center of Quantum Matter, Beijing, China}
\affiliation{Center for High Energy Physics, Peking University, Beijing 100871, China}

\date{\today}



\begin{abstract}
We calculate $\Lambda$ and $\bar{\Lambda}$ hadron production cross sections in charged lepton semi-inclusive deep-inelastic scattering off nuclear target ($A$, using iron $\mathrm{Fe}$ as an example) and deuteron ($D$) target. The results show that the ratio $(\bar{\Lambda}^{A}/\Lambda^{A})/(\bar{\Lambda}^{D}/\Lambda^{D})$ is sensitive to the sea quark content of the nucleus. We adopt three different models to take the nuclear EMC effect into account. The ratio $(\bar{\Lambda}^{A}/\Lambda^{A})/(\bar{\Lambda}^{D}/\Lambda^{D})$ is predicted to be different by these different models.
\end{abstract}

\maketitle



\section{Introduction}\label{sec:intro}
The European Muon Collaboration (EMC) found in 1983 that the ratio of structure functions per nucleon of iron (Fe) to deuterium (D) is different from earlier prediction by taking into account the Fermi motion of bound nucleons~\cite{Aubert:1983xm}. This is the so called nuclear EMC effect, which implies that the quark structure of a bound nucleon is different from that of a free one. Since the discovery of the EMC effect, lots of experimental measurements have been performed in charged lepton-nucleus
scattering~\cite{Arnold:1983mw,Aubert:1986yn,Dasu:1988ru,Dasu:1993vk,Gomez:1993ri,Seely:2009gt}, neutrino-nucleus scattering~\cite{Abramowicz:1984yk,Parker:1983yi,CooperSarkar:1984eb,Ammosov:1984rd,Hanlon:1985yg,Guy:1986us}, and the Drell-Yan process~\cite{Close:1984zn,McGaughey:1999mq,Chmaj:1983jq,Chmaj:1985pp,DiasdeDeus:1984wy,DiasdeDeus:1984ge}.
Many theoretical and phenomenological models were proposed to describe the data in the intermediate $x$ region for the nuclear EMC effect, such as the pion excess model~\cite{LlewellynSmith:1983vzz,Ericson:1983um,Berger:1983jk}, the quark-cluster model~\cite{Pirner:1980eu,Jaffe:1982rr,Carlson:1983fs}, and the rescaling model~\cite{Jaffe:1982rr,Close:1983tn,Jaffe:1983zw,Close:1984zn,Nachtmann:1983py}. However, these models provide totally different pictures about the nuclear structure. The pion excess model supposes that a nucleus contain hadronic constituents (mostly pions) other than nucleons, naturally predicting an increase of $\bar{u}$ and $\bar{d}$ sea quarks in the nucleus compared to those in a free nucleon. A nucleus is assumed to be composed of nucleons and multiquark clusters (in which more than three quarks are confined together) in the quark-cluster model. In the rescaling model, the quark confinement size of a nucleon in the nucleus is assumed to be bigger than that in free nucleons~\cite{Jaffe:1982rr}.

Though different models can explain the measured ratio of structure functions of bound nucleons to free ones, they provide quite different predictions of the sea quark content of the nucleus. The modification of the sea quark distributions in the nucleus was considered to match the gap between the theoretical calculations and the experimental data from many scattering processes. In fact, the ratio of bound structure functions to free ones through inclusive deep inelastic scattering (DIS) processes is actually insensitive to the nuclear sea content. Then the experimental processes and quantities that are sensitive to the sea content in the nucleus should enable us to discriminate different models. It was shown that the $\Lambda/\bar{\Lambda}$ ratio of production cross sections in semi-inclusive DIS (SIDIS) process is a physical quantity that is sensitive to the sea quark content of nucleons~\cite{Ma:2004zt,Lu:2006xr,Chi:2013hka,Chi:2014xba,Du:2017nzy}. In Ref.~\cite{Lu:2006xr}, the hadron production ratios of $(\bar{\Lambda}^{A}/\Lambda^{A})/(\bar{\Lambda}^{D}/\Lambda^{D})$ in the SIDIS process are found to be quite different in the pion excess model, the quark-cluster model, and the rescaling model. In our present work, we reconsider these three models by fitting the experimental data of the EMC ratio from the EMC Collaboration~\cite{Aubert:1983xm} and from the BCDMS Collaboration~\cite{Benvenuti:1987az} with the uncertainties considered. We significantly improve the numerical analysis, based on the new development of the $\Lambda$ and $\bar{\Lambda}$ fragmentation functions~\cite{Du:2017nzy} and more careful analysis of the nuclear models for the EMC effect.

In Sec.~\ref{sec:ww1}, we make a detailed description of three different nuclear models for the EMC effect, namely the pion excess model, the quark-cluster model, and the rescaling model. The detailed modification of $\Lambda$ and $\bar{\Lambda}$ fragmentation functions is discussed in Sec.~\ref{sec:ww2}. In Sec.~\ref{sec:ww3}, we make an analysis of different production ratios of $\Lambda$ and $\bar{\Lambda}$ in charged lepton semi-inclusive deep-inelastic scattering off a nuclear target (iron Fe) and a deuteron (D) target. Our work shows that the effects of experimental errors on the results corresponding to the pion excess model and the rescaling model are small. Though the error bands corresponding to the quark-cluster model are wide, the ratio $(\bar{\Lambda}^{A}/\Lambda^{A})/(\bar{\Lambda}^{D}/\Lambda^{D})$ can still enable us to discriminate these three models of the EMC effect to some extent. We show that the physical quantity $(\bar{\Lambda}^{A}/\Lambda^{A})/(\bar{\Lambda}^{D}/\Lambda^{D})$ is a good window to figure out new features of the EMC effect.

\section{Three Nuclear Models For THE EMC Effect}\label{sec:ww1}

\begin{figure}
	\begin{center}
		\includegraphics[scale=0.3]{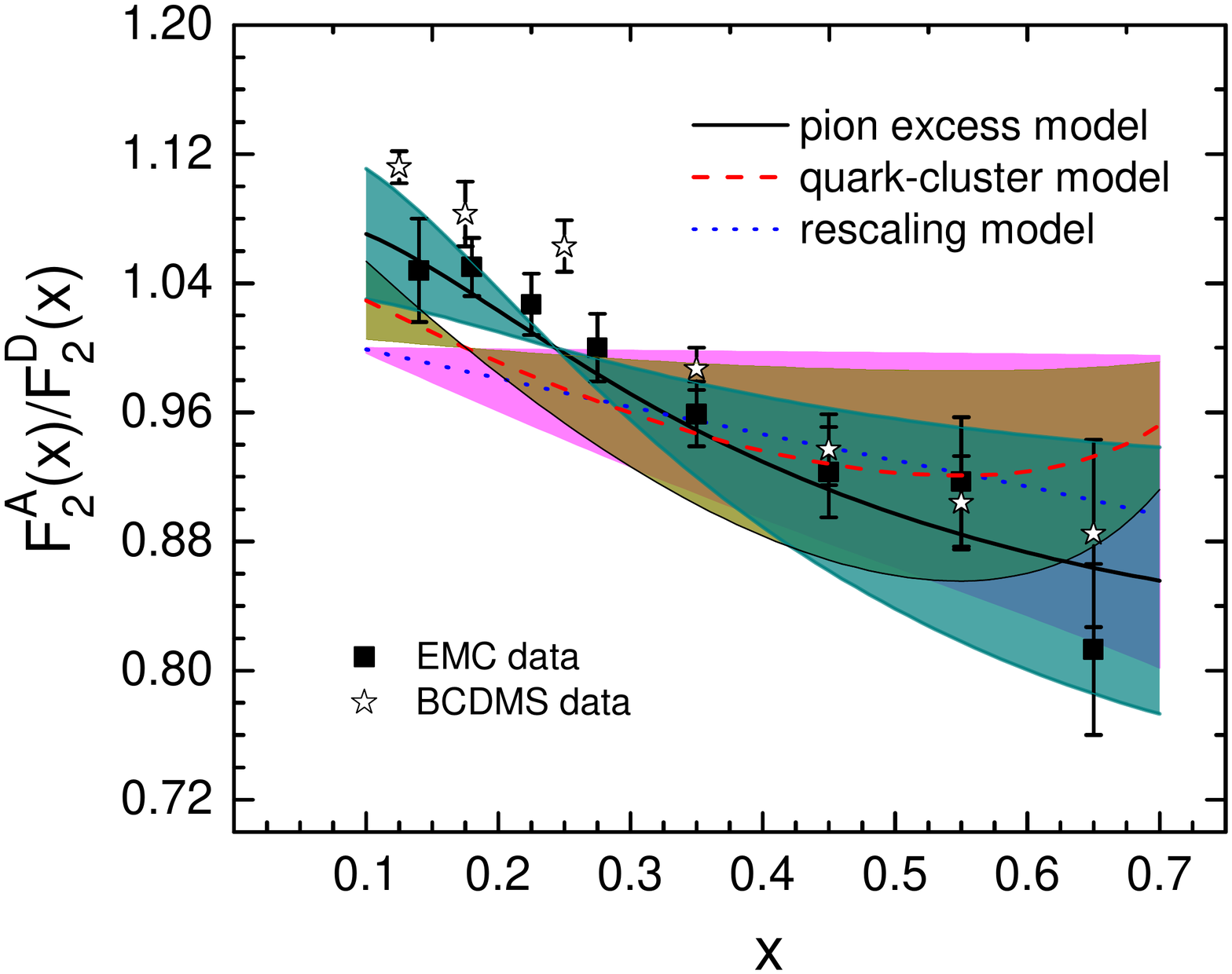}
	\end{center}
	\vspace{-0.5cm}
	\caption{\label{fig:20} The results of $F_{2}^{A}/F_{2}^{D}$ at $Q^{2}=5~\mathrm{GeV^{2}}$. The solid-black, dashed-red, and dotted-blue lines are the results of the pion excess model, the quark-cluster model, and the rescaling model. The nucleus $A$ is Fe.}
\end{figure}

\begin{figure}
	\begin{center}
		\includegraphics[scale=0.3]{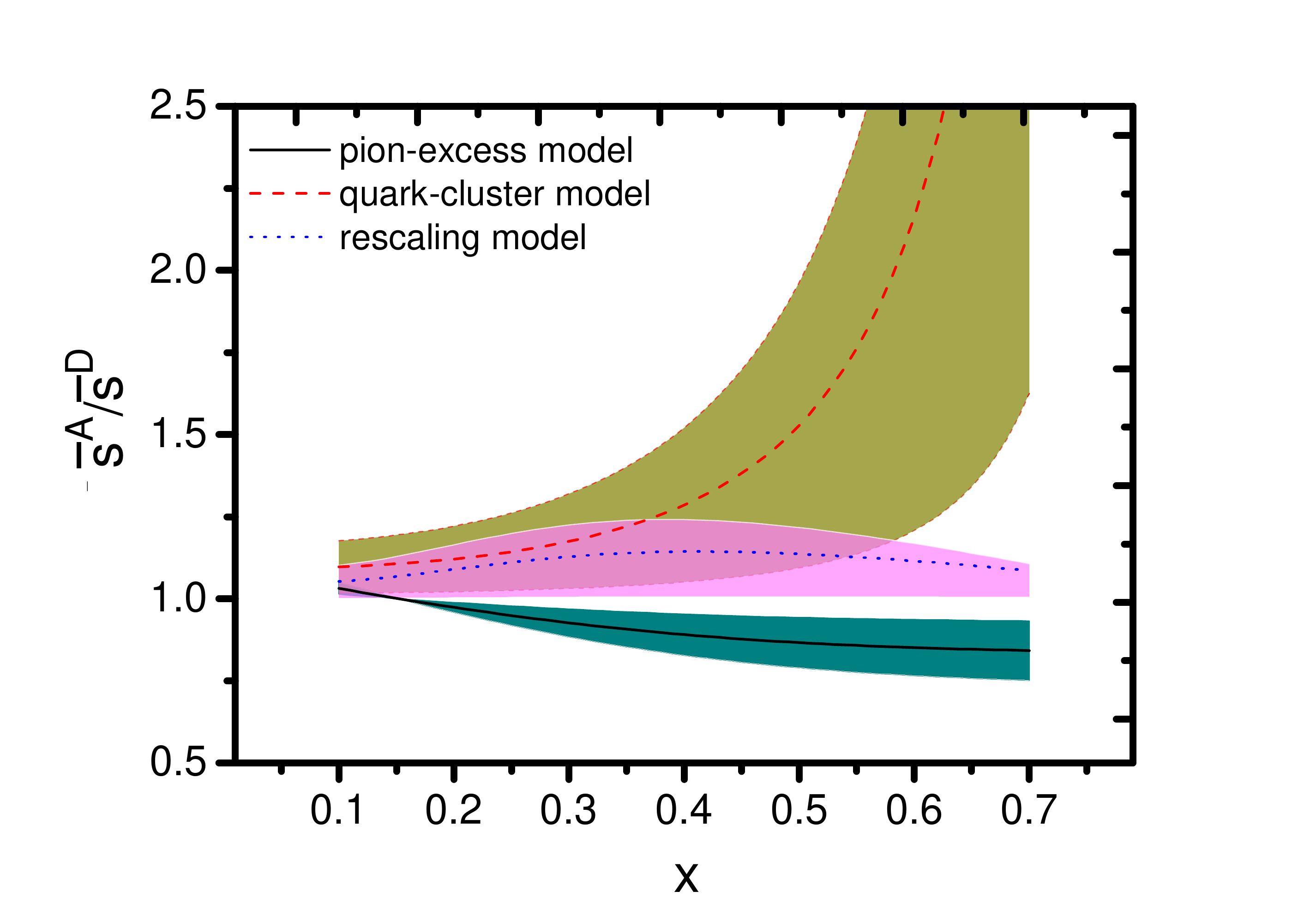}
	\end{center}
	\vspace{-0.5cm}
	\caption{\label{fig:22} The results of $\bar{s}^{A}/\bar{s}^{D}$. The solid-black, dashed-red, and dotted-blue lines are the results of the pion excess model, the quark-cluster model, and the rescaling model. The nucleus $A$ is Fe.}
\end{figure}

The EMC effect is not consistent with the prediction that the cross section for muon scattering on a nucleus should be almost the sum of the cross sections of free nucleons in the nucleus. The SLAC E139 Collaboration released precise data of the cross section ratio $\sigma^{A}/\sigma^{D}$ for several nuclei in the region $x>0.2$~\cite{Arnold:1983mw}, with $\sigma^{A}$ and  $\sigma^{D}$ being the per-nucleon cross sections in nucleus $A$ and deuteron $D$ respectively. The E139 experiment observed a reduction of the ratio $\sigma^{A}/\sigma^{D}$ in the region of $0.3<x<0.8$ for all measured nuclei. The JLab-E03103 Collaboration also released data for $\sigma^{C(N)}/\sigma^{D}$ in the region $x>0.3$~\cite{Seely:2009gt}. All these data show that, in the intermediate $x$ region, the structure function in the nucleus is smaller than that in a free nucleon.
In the inclusive deep inelastic scattering (DIS) process, the cross section ratio $\sigma^{A}/\sigma^{D}$ can be expressed as $F_{2}^{A}(x,Q^{2})/F_{2}^{D}(x,Q^{2})$. In the naive quark model, the ratio $F_{2}^{A}(x,Q^{2})/F_{2}^{D}(x,Q^{2})$ is written as

\begin{equation}
\frac{F_{2}^{A}(x,Q^{2})}{F_{2}^{D}(x,Q^{2})}=\frac{\sum_{i}e_{i}^{2}[q_{i}(x,Q^{2},A)+\bar{q}_{i}(x,Q^{2},A)]}{\sum_{i}e_{i}^{2}[q_{i}(x,Q^{2})+\bar{q}_{i}(x,Q^{2})]},
\end{equation}
where $e_{i}$ means the charge of a parton with flavor $i$, and $q_{i}(x,Q^{2})$ are the parton distribution functions (PDFs) of a nucleon in deuteron, with $q_{i}(x,Q^{2},A)$ being the PDFs per-nucleon in nucleus $A$. In Fig.~\ref{fig:20}, we draw the ratios $F_{2}^{A}/F_{2}^{D}$ of the pion excess model, the quark-cluster model, and the rescaling model at $Q^{2}=5~\mathrm{GeV^{2}}$ in the region $0.1<x<0.7$. All these three models reproduce a reduction of $F_{2}^{A}/F_{2}^{D}$ in the intermediate $x$ region, roughly being consistent with the phenomena of the nuclear EMC effect. However, the large different predictions of sea quark distributions for these three models are illustrated to discriminate these three models, as shown in Figs.~\ref{fig:21} and ~\ref{fig:22}. We notice that the sea content in nuclei for the quark-cluster model is much more enhanced than that in nucleons.

\subsection{The pion excess model}
It was first proposed by Llewellyn Smith that the behavior of $F_{2}(Fe)/F_{2}(D)$ in the EMC effect could be accounted for by the excess number of pions per nucleon~\cite{LlewellynSmith:1983vzz}. Later, the quantitative expression of how extra pions contribute to the nuclear structure function was proposed in~\cite{Ericson:1983um}:
\begin{equation}
  \delta F_{2}^{N}(x,Q^{2})=\int_{x}^{1}f_{\pi}^{A}(y)F_{2}^{\pi}(x/y,Q^{2}) \d y,
\end{equation}
where $F_{2}^{\pi}$ is the structure function of the pion, and $x$ is the momentum fraction of the struck quark in the nucleon-pion subsystem, and $f_{\pi}^{A}(y)$ is the probability of finding pions carrying a fraction $y$ of the momentum of the nucleon-pion subsystem in nucleus $A$. $f_{\pi}^{A}(y)$ satisfies the normalization
\begin{equation}
\int_{0}^{1}f_{\pi}^{A}(y)\d y=n_{\pi},
\end{equation}
where $n_{\pi}$ is the number of pions per nucleon. So, the form of the nuclear structure function can be expressed as
\begin{equation}
  F_{2}^{A}=\int_{x}^{1}f_{\pi}^{A}(y)F_{2}^{\pi}(x/y,Q^{2})\d y+\int_{x}^{1}f_{N}^{A}(z)F_{2}^{N}(x/z,Q^{2})\d z,
\end{equation}
where $F_{2}^{\pi}$ and $F_{2}^{N}$ are structure functions of free nucleons and free pions, with $f_{N}^{A}(z)$ being the number density of nucleons per nucleon in nucleus $A$, and $z$ being the nucleon momentum fraction of the nucleon-pion subsystem. $f_{N}^{A}(z)$ meets the normalization $\int f_{N}^{A}(z)\d z=1$. Then in the naive parton model, the quark distribution of the nucleus can be written as
\begin{equation}
q_{i}^{A}=\int_{x}^{1}\frac{\d y}{y}f_{\pi}^{A}(y)q_{i}^{\pi}(\frac{x}{y})+\int_{x}^{1}\frac{\d z}{z}f_{N}^{A}(z)q_{i}^{N}(\frac{x}{z}),
\end{equation}
in which $q_{i}^{\pi}$ and $q_{i}^{N}$ are the parton distribution functions of the free pion and free nucleon. We adopt a toy model parametrization~\cite{Berger:1983jk}, which is written as
\begin{equation}
  f_{\pi}^{A}(y)=\left<n_{\pi}\right>\frac{\Gamma(a+b+2)}{\Gamma{(a+1)}\Gamma{(b+1)}}y^{a}(1-y)^{b}.
\end{equation}

The selection of parameters $a$ and $b$ may be varied so as to adjust the $y$ dependence of $f_{\pi}^{A}(y)$, here we choose $a=1$ and $b=3$~\cite{Berger:1983jk}. In this toy model, the bound nuclei are assumed to be combinations of free nucleons and nucleon-pion subsystems. The exchanged virtual photon in the DIS process may interact with free nucleons in nuclei as $z$ approaches to $1$. The longitudinal momentum fraction of the nucleon in the nucleon-pion subsystem is $z$ while the probability of finding a pion carrying a momentum fraction $y=1-z$ of the subsystem is $f_{\pi}^{A}(y)$. So we know that
\begin{equation}
f_{N}^{A}(z)=(1-\left<n_{\pi}\right>)\delta(z-1)+f_{\pi}^{A}(1-z).
\end{equation}

\begin{figure}
	\begin{center}
		\includegraphics[scale=0.3]{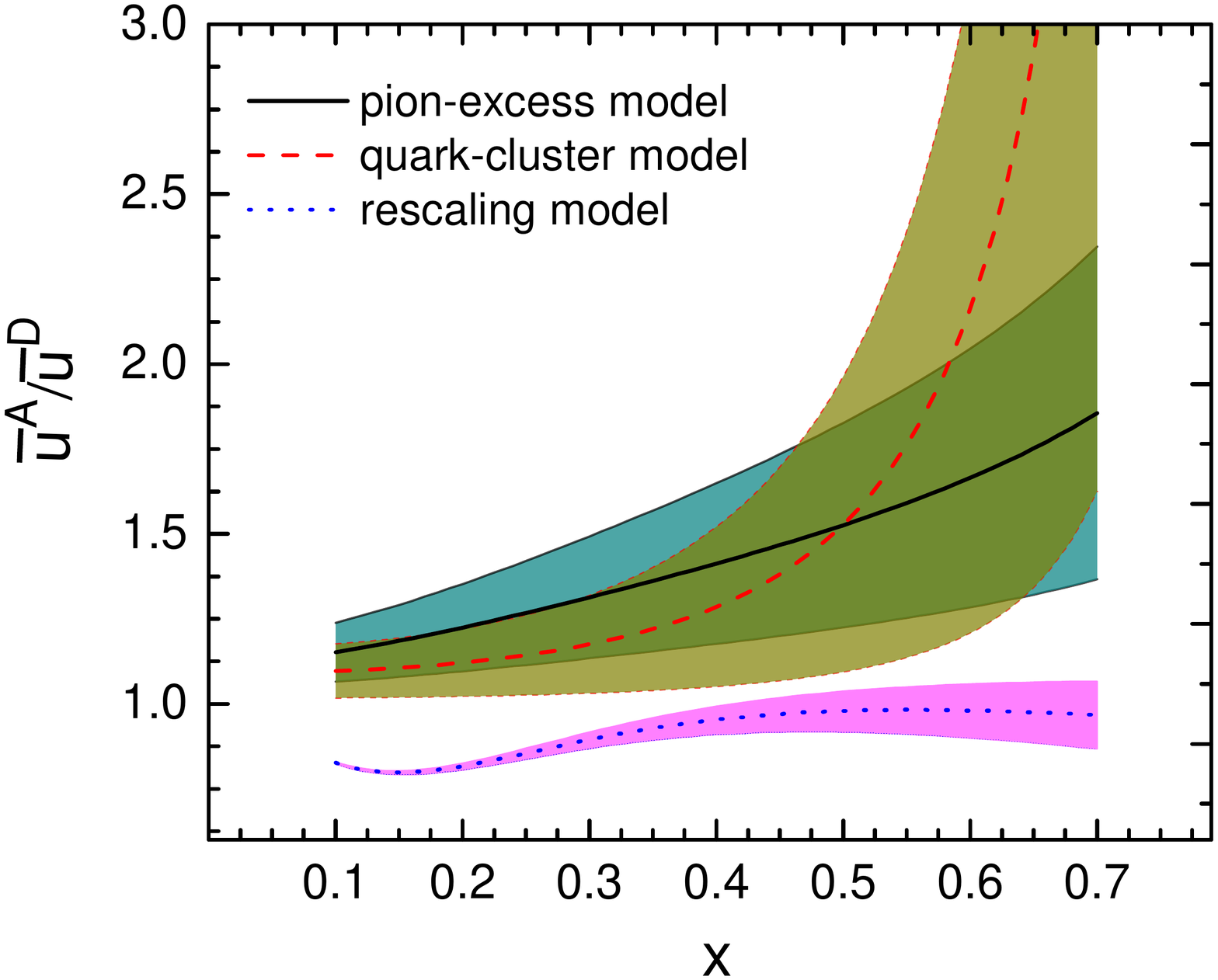}
	\end{center}
	\vspace{-0.5cm}
	\caption{\label{fig:21} The results of $\bar{u}^{A}/\bar{u}^{D}$. The solid-black, dashed-red, and dotted-blue lines are the results of the pion excess model, the quark-cluster model, and the rescaling model. The nucleus $A$ is Fe.}
\end{figure}

The parton distributions of the nucleon are from the CTEQ14L parametrization~\cite{Dulat:2015mca}. For the parton distributions of free pions, we adopt the MRS parametrization~\cite{Sutton:1991ay}. To define the value of the parameter $\left<n_{\pi}\right>$ in the pion excess model, we fit the experimental data of the EMC ratio $F_{2}^{A}/F_{2}^{D}$ using MINUIT~\cite{James:1975dr}. The fitting results are shown in Table~\ref{tab:fitting results}, and we also consider the errors of the experimental data. From the calculations, we notice that the pion excess model can roughly describe the behavior of $F_{2}^{A}/F_{2}^{D}$ in the intermediate $x$ region. What is more, the pion excess model predicts an enhancement of $u$ sea distributions in the nucleus, as shown in Fig.~\ref{fig:21}.

\begin{widetext}
\begin{center}
\begin{table}
\begin{tabular}{|c|c|c|c|}\hline
Models & Experimental data & Fitting parameters & Fitting results \\ \hline
pion excess model  &  EMC~\cite{Aubert:1983xm}+BCDMS~\cite{Benvenuti:1987az} & $\left<n_{\pi}\right>$  &  $0.166 \pm0.095$ \\ \hline
quark-cluster model  &  EMC~\cite{Aubert:1983xm}+BCDMS~\cite{Benvenuti:1987az}  & $f$   &  $0.2524 \pm0.2079$ \\ \hline
rescaling model  &    EMC~\cite{Aubert:1983xm}+BCDMS~\cite{Benvenuti:1987az}    & $\xi_{A}$ &   $1.5475^{+0.959}_{-0.529}$ \\ \hline
\end{tabular}
\caption{\label{tab:fitting results} The fitting results are $\left<n_{\pi}\right>$ in the pion excess model, $f$ in the quark-cluster model, and $\xi_{A}$ in the rescaling model. The nucleus $A$ is Fe.}
\end{table}
\end{center}
\end{widetext}

\subsection{The quark-cluster model}
As we know, nucleons are tightly bound together in the nucleus. The quark-cluster model assumes that there exists a chance of more than three quarks to be confine together in a nucleus.

As there are no experimental data about the distributions of six-quark clusters, Ref.~\cite{Carlson:1983fs} adopts counting rules~\cite{Blankenbecler:1974tm,Farrar:1975yb,Vainshtein:1977db,Sivers:1982wk}, expressing the parton distributions of the proton as
\begin{eqnarray}
  U_{v}(x)&=&xu_{v}(x)=2N_{u}\sqrt{x}(1-x)^{3},\nonumber \\
  D_{v}(x)&=&xd_{v}(x)=N_{d}\sqrt{x}(1-x)^{4},\nonumber \\
  \bar{U}(x)&=&x\bar{u}(x)=\bar{N}(1-x)^{7},
\end{eqnarray}
and $N_{u}$, $N_{d}$, and $\bar{N}$ need to match the following normalizations:
\begin{gather}
\int_{0}^{1}u_{v}(x)\d x=2, \\
\int_{0}^{1}d_{v}(x)\d x=1, \\
\left<x_{\mathrm{sea}}\right>=1-\left<x_{\mathrm{valence}}\right>-\left<x_{\mathrm{gluon}}\right>,
\end{gather}
where $\left<x_{\mathrm{gluon}}\right>$ is the longitudinal momentum fraction carried by gluons. We take $\left<x_{\mathrm{gluon}}\right>=0.57$ as suggested by experimental results~\cite{deGroot:1978feq}. From calculations of Eqs.~(9)-(11), we get $N_{u}=1.094$, $N_{d}=1.2305$, and $\bar{N}=0.1869$.

The shapes of sea quark distributions are assumed to be~\cite{Abramowicz:1982zr}
\begin{equation}
s(x)=\frac{1}{2}\bar{u}(x)=\frac{1}{2}\bar{d}(x).
\end{equation}
So the structure function of the proton is
\begin{equation}
F_{2}^{p}(x,Q^{2})=x~\sum_{i}e_{i}^{2}[q_{i}(x)+\bar{q}_{i}(x)], ~~~~    (i=u ~\mathrm{or}~ d).
\end{equation}
For the structure function of the neutron, $F_{2}^{n}$, we apply the isospin asymmetry between proton and neutron.

The parton distributions of a six-quark cluster can be written as~\cite{Carlson:1983fs},
\begin{eqnarray}
  &V_{6}=zv_{6}(z)=3N_{6}\sqrt{z}(1-z)^{10}, \\
  &\bar{U}_{6}(z)=z\bar{u}_{6}(z)=\bar{N}_{6}(1-z)^{14},
\end{eqnarray}
where $N_{6}$ and $\bar{N}_{6}$ match the following normalizations:
\begin{eqnarray}
\int_{0}^{1}v_{6}(z)\d z&=&3, \\
\left<z_{\mathrm{sea}}\right>&=&1-\left<z_{\mathrm{valence}}\right>-\left<z_{\mathrm{gluon}}\right>.
\end{eqnarray}

It turns out that $N_{6}=1.850$ and $\bar{N}_{6}=0.5042$. Here $z=x/2$, with $x$ being the momentum fraction of one single nucleon. Ref.~\cite{Carlson:1983fs} assumes that the probability of finding six-quark clusters in the nucleus is $f$, and then
\begin{eqnarray}
  q^{A}(x)&=&(1-f)q^{N}(x)+f\frac{q^{6}(\frac{x}{2})}{4},\\
  \frac{F_{2}^{A}(x,Q^{2})}{F_{2}^{D}(x,Q^{2})}&=&(1-f)+f\frac{F_{2}^{6}(x,Q^{2})}{F_{2}^{D}(x,Q^{2})},
\end{eqnarray}
where
\begin{eqnarray}
 F_{2}^{6}(x,Q^{2})&=&x~\sum_{i}e_{i}^{2}[\frac{q_{i}^{6}(\frac{x}{2})}{4}+\frac{\bar{q}_{i}^{6}(\frac{x}{2})}{4}]\nonumber \\
                   &=&\frac{5}{18}V^{6}(\frac{x}{2})+\frac{11}{18}\bar{U}_{6}(\frac{x}{2}),\\
F_{2}^{D}(x,Q^{2})&=&\frac{1}{2}[F_{2}^{p}(x,Q^{2})+F_{2}^{n}(x,Q^{2})]\nonumber \\
                  &=&\frac{5}{18}[U_{v}(x)+D_{v}(x)]+\frac{11}{9}\bar{U}(x).
\end{eqnarray}
Here, the deuteron is considered without including the EMC effect: it is assumed to be just a combination of one proton and one neutron. What is more, the isospin symmetry between proton and neutron is adopted. We notice that Eq.~(18) satisfies the correct counting $\int_{0}^{2}q^{A}_{v}(x)\mathrm{d} x=3/2$ for each valence quark per nucleon for a nucleus with six-quark clusters.  

The value of the parameter $f$ in our work is given by fitting the experimental data. The results are shown in Table~\ref{tab:fitting results}. Also, we plot the EMC ratio for the quark-cluster model in Fig.~\ref{fig:20}.

\subsection{The rescaling model}
It was observed by Close, Roberts, and Ross~\cite{Close:1983tn} that the nuclear structure function of the nucleus roughly equals to that of the deuteron at a higher value of $Q^{2}$,
\begin{eqnarray}
  F_{2}^{A}(x,Q^{2})&=&F_{2}^{N}(x,\xi_{A}(Q^{2})Q^{2}),\\
  q^{A}(x,Q^{2})&=&q^{N}(x,\xi_{A}(Q^{2})Q^{2}),
\end{eqnarray}
where $\xi_{A}$ is the rescaling factor at $Q^{2}$. This phenomenon is interpreted as being due to the different quark confinement sizes between the deuteron and the iron~\cite{Jaffe:1983zw}. The CTEQ14L parametrization~\cite{Dulat:2015mca} is adopted for the parton distribution function of a free nucleon. To define the value of $\xi_{\mathrm{Fe}}$, we fit the experimental data of the EMC ratio for iron. The fitting results are shown in Table~\ref{tab:fitting results}.

\section{semi-inclusive deep inelastic scattering}\label{sec:ww2}
For semi-inclusive production of hadron $h$ in the deep inelastic scattering process, the cross section ratio can be expressed as
\begin{widetext}
\begin{equation}
\frac{d\sigma_{A}^{h}/dx}{d\sigma_{D}^{h}/dx}=
\frac{\int_{a}^{b}dz\sum_{i}e_{i}^{2}[q_{i}^{A}(x,Q^{2})D_{q_{i}}^{h}(z,Q^{2},A)+\bar{q}_{i}^{A}(x,Q^{2})D_{\bar{q}_{i}}^{h}(z,Q^{2},A)]}
{\int_{a}^{b}dz\sum_{i}e_{i}^{2}[q_{i}^{D}(x,Q^{2})D_{q_{i}}^{h}(z,Q^{2})+\bar{q}_{i}^{D}(x,Q^{2})D_{\bar{q}_{i}}^{h}(z,Q^{2})]},
\end{equation}
\end{widetext}
where $q_{i}(x,Q^{2})$ is the parton distribution of quark $q_{i}$ with flavor $i$, and $D_{q_{i}}^{h}(x,Q^{2})$ means the fragmentation function for quark ${q_{i}}$ to hadron $h$ in free nucleons. We label the fragmentation function in the nucleus as  $D_{q_{i}}^{h}(x,Q^{2},A)$. During our calculations, the parton distribution of the deuteron is treated as the averaged value of a proton and a neutron. Besides, as we mentioned early in the paper, the three models for the EMC effect predict largely different sea quark distributions in the nucleus. So by adopting different models, this ratio can reveal how the EMC effect depends on sea quark distributions in the nucleus.

Due to the nonperturbative nature of the fragmentation process, we need to obtain the fragmentation functions from some phenomenological parametrizations. The Gribov-Lipatov relation~\cite{Gribov:1971zn,Barone:2000tx} is adopted,
\begin{equation}
  D_{q}^{h}(z)\propto q^{h}(z),
\end{equation}
where $D_{q}^{h}(z)$ denotes the fragmentation function of a quark $q$ to a hadron $h$ with momentum fraction $z$ and $q^{h}(z)$ denotes
the quark distribution function at momentum fraction $z$ inside a hadron $h$.
To distinguish between the valence part and the sea part, we write the relation in a detailed way~\cite{Ma:2003gd},
\begin{eqnarray}
  D_{V}^{h}(z)&=&C_{V}z^{\alpha}q_{V}^{h}(z),\\
  D_{S}^{h}(z)&=&C_{S}z^{\alpha}q_{S}^{h}(z),
\end{eqnarray}
where $D_{S}^{h}(z)$ means the fragmentation function of hadron $h$ from sea quarks in the nucleon, and $D_{V}^{h}(z)$ means the fragmentation function from the valence part in the nucleon. So the total fragmentation functions for hadron $\Lambda$ are written as~\cite{Ma:2003gd}\
\begin{eqnarray}
  D_{q}^{\Lambda}&=&D_{V}^{\Lambda}+D_{S}^{\Lambda},\\
  D_{\bar{q}}^{\Lambda}&=&D_{S}^{\Lambda}.
\end{eqnarray}
We can calculate different sets of $C_{V}$, $C_{S}$, and $\alpha$ to show how the sea part in a nucleus influences the results. Three sets are adopted~\cite{Ma:2003gd}: (1) $C_{V}=1$ and $C_{S}=0$ for $\alpha=0$; (2) $C_{V}=1$ and $C_{S}=1$ for $\alpha=0.5$; (3) $C_{V}=1$ and $C_{S}=3$ for $\alpha=1$. From the Gribov-Lipatov relation, we see that the parton distributions of hadron $\Lambda$ are also needed to obtain the fragmentation functions. In our paper, the SU(3) symmetry~\cite{Ma:2001ri} is used to get the $\Lambda$ parton distributions from proton parton distribution functions, which are from the CTEQ14L parametrization~\cite{Dulat:2015mca}. Here we need to say that the differences of $\Lambda$ parton distributions between SU(3) symmetry model and other models are small in the intermediate $x$ region~\cite{Ma:1999gj}. Then it is reasonable for us to show how other factors affect the final results.

In our work, we calculate the production ratio of hadrons $\Lambda$ and $\bar{\Lambda}$, shown in Figs.~\ref{fig:23}, ~\ref{fig:24}, and ~\ref{fig:25}. Three different models and three sets of fragmentation functions are discussed. What is more, we use the improved parametrization in Ref.~\cite{Du:2017nzy} for fragmentation functions of $\Lambda$ and $\bar{\Lambda}$ from free nucleons to adjust our numerical calculations. The corresponding results are shown in Figs.~\ref{fig:26}, ~\ref{fig:27}, and ~\ref{fig:28}. This parametrization~\cite{Du:2017nzy} takes into account the enhancement of strange quark to $\Lambda$ productions more naturally and can give more reasonable results:
\begin{eqnarray}
D_{u}^{\Lambda}(z,Q^{2})&=&(\frac{D_{u}^{\Lambda}}{D_{u+\bar{u}}^{\Lambda}})^{\mathrm{th}}D_{u+\bar{u}}^{\Lambda}(z,Q^{2})^{\mathrm{AKK}},\nonumber \\
D_{\bar{u}}^{\Lambda}(z,Q^{2})&=&(\frac{D_{\bar{u}}^{\Lambda}}{D_{u+\bar{u}}^{\Lambda}})^{\mathrm{th}}D_{u+\bar{u}}^{\Lambda}(z,Q^{2})^{\mathrm{AKK}},\nonumber \\
D_{d}^{\Lambda}(z,Q^{2})&=&(\frac{D_{d}^{\Lambda}}{D_{d+\bar{d}}^{\Lambda}})^{\mathrm{th}}D_{d+\bar{d}}^{\Lambda}(z,Q^{2})^{\mathrm{AKK}},\nonumber \\
D_{\bar{d}}^{\Lambda}(z,Q^{2})&=&(\frac{D_{\bar{d}}^{\Lambda}}{D_{d+\bar{d}}^{\Lambda}})^{\mathrm{th}}D_{d+\bar{d}}^{\Lambda}(z,Q^{2})^{\mathrm{AKK}},\nonumber \\
D_{s}^{\Lambda}(z,Q^{2})&=&(\frac{D_{s}^{\Lambda}}{D_{u+\bar{s}}^{\Lambda}})^{\mathrm{th}}D_{s+\bar{s}}^{\Lambda}(z,Q^{2})^{\mathrm{AKK}},\nonumber \\
D_{\bar{s}}^{\Lambda}(z,Q^{2})&=&(\frac{D_{\bar{s}}^{\Lambda}}{D_{s+\bar{s}}^{\Lambda}})^{\mathrm{th}}D_{s+\bar{s}}^{\Lambda}(z,Q^{2})^{\mathrm{AKK}},
\end{eqnarray}
where the quantities with superscripts $\mathrm{AKK}$ denote the Albino, Kniehl, and Kramer (AKK) parametrization of quark to $\Lambda$ fragmentation functions~\cite{Albino:2008fy}.

Then we need to distinguish between the fragmentation functions in the nucleus and those in the free nucleon. The HERMES collaboration found that the hadron production cross section from the nucleus is lower than that from the free nucleon~\cite{Airapetian:2003mi}. Many mechanisms were proposed to explain this HERMES data, such as nuclear absorption of the produced hadron~\cite{Bialas:1983kn,Bialas:1986cf}, partial deconfinement in the nucleus~\cite{Accardi:2002tv,Jaffe:1983zw}, and the energy-loss model~\cite{Wang:1996yh,Wang:2002ri,Arleo:2003jz}. In this work, we adopt the energy-loss model to calculate the fragmentation functions in the nucleus.

In the energy-loss model~\cite{Arleo:2003jz}, due to the existence of nuclear medium, the quark energy is reduced from $E=\nu$ to $E=\nu-\varepsilon$ during the hadronization process. Then the momentum fraction of hadron production is modified,
\begin{equation}
  z=\frac{E_{h}}{\nu}  \rightarrow  z^{*}=\frac{E_{h}}{\nu-\varepsilon}=\frac{z}{1-\varepsilon/\nu},
\end{equation}
where $E_{h}=\nu-\varepsilon$ is the measured hadron energy and $\varepsilon$ is the loss energy of the quark during the hadronization. The nucleus fragmentation function is expressed as
\begin{equation}
  zD_{q}^{h}(z,Q^{2},A)=\int_{0}^{\nu-E_{h}}\d \varepsilon D(\varepsilon, \nu)z^{*}D_{q}^{h}(z^{*},Q^{2}),
\end{equation}
in which $D(\varepsilon,\nu)$ is the probability for a quark with energy $\nu$ to lose an energy $\varepsilon$. Here we adopt the parametrization by Arleo~\cite{Arleo:2002kh},
\begin{equation}
  D(\varepsilon)=\frac{1}{\sqrt{2\pi}\sigma\varepsilon}\exp[-\frac{(\ln(\varepsilon/\omega_{c})-\nu)^{2}}{2\sigma^{2}}],
\end{equation}
where $\mu$ and $\sigma$ are parametrized as $\mu=-1.5$ and $\sigma=0.73$, and $\omega_{c}$ indicates the energy loss scale of the fragmented quark,
\begin{equation}
  \omega_{c}=\frac{1}{2}\hat{q}L^{2}.
\end{equation}
Here the so-called gluon transport coefficient $\hat{q}$ measures the scattering power of the nuclear medium. We choose the set $\hat{q}=0.72~\mathrm{GeV/fm^{2}}$ and $L=3R/4$ in Ref.~\cite{Arleo:2002kh}. $R\cong1.17A^{1/3}~\mathrm{fm}$ denotes the nuclear radius of nucleus $A$. For a Fe nucleus, we set $\nu=12~\mathrm{GeV}$. However, the probability function $D(\varepsilon,\nu)$ has a different normalization from the probability function $D(\varepsilon)$. So we need to transform the probability function in the following way:
\begin{eqnarray}
  D(\varepsilon,\nu)=n_{\nu}\frac{1}{\sqrt{2\pi}\sigma\varepsilon}\exp[-\frac{(\ln(\varepsilon/\omega_{c})-\nu)^{2}}{2\sigma^{2}}],
\end{eqnarray}
where $n_{\nu}$ is the normalization coefficient through the condition
\begin{eqnarray}
  \int_{0}^{\nu-E_{h}}d\varepsilon D(\varepsilon,\nu)=1.
\end{eqnarray}
The different normalization factors reflect the difference between our work and Ref.~\cite{Lu:2006xr}.

\section{Results}\label{sec:ww3}

\begin{figure*}[!h]
	\begin{center}
		\includegraphics[scale=0.22]{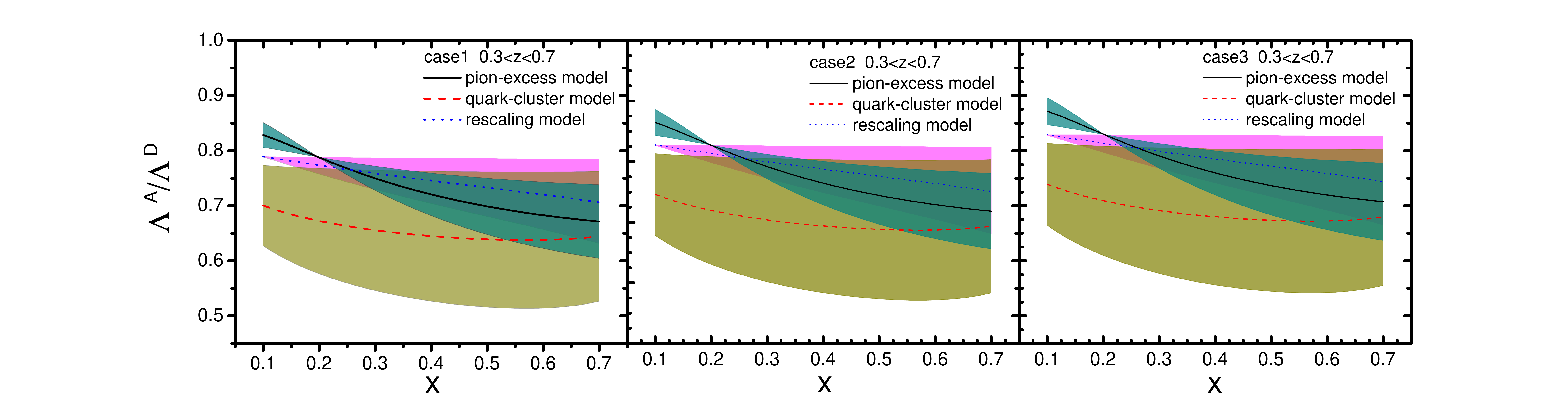}
	\end{center}
	\vspace{-0.5cm}
	\caption{\label{fig:23} The results of $\Lambda^{A}/\Lambda^{D}$ at $Q^{2}=5~\mathrm{GeV^{2}}$. The solid-black, dashed-red, and dotted-blue curves are the results of the pion excess model, the quark-cluster model, and the rescaling model. The nucleus $A$ is Fe. These three figures show different fragmentation function cases.}
\end{figure*}

\begin{figure*}
	\begin{center}
		\includegraphics[scale=0.22]{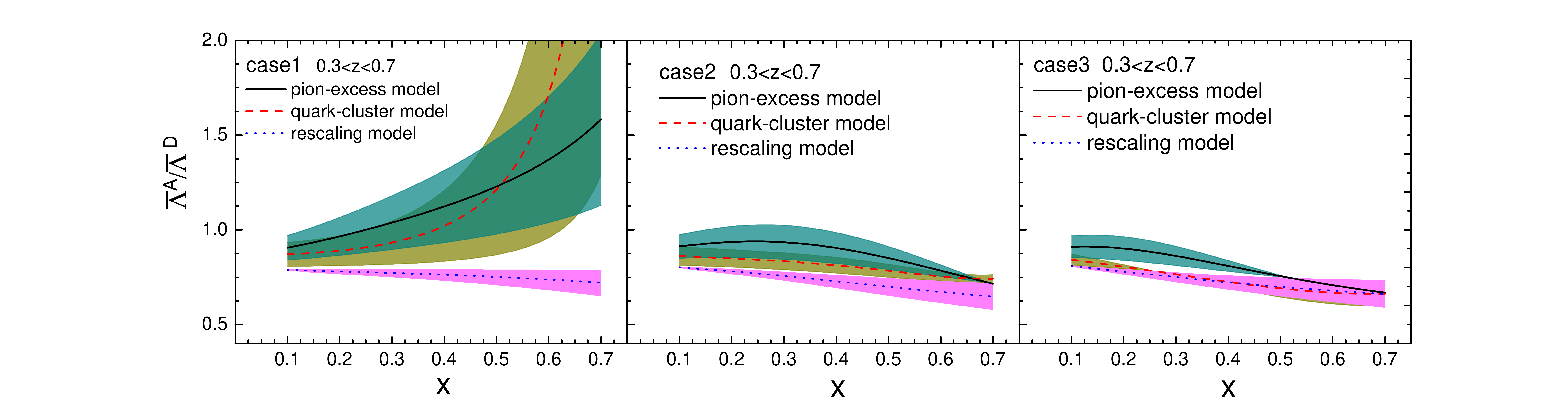}
	\end{center}
	\vspace{-0.5cm}
	\caption{\label{fig:24} The results of $\bar{\Lambda}^{A}/\bar{\Lambda}^{D}$ at $Q^{2}=5~\mathrm{GeV^{2}}$. The solid-black, dashed-red, and dotted-blue curves are the results of the pion excess model, the quark-cluster model, and the rescaling model. The nucleus $A$ is Fe. These three figures show different fragmentation function cases.}
\end{figure*}

\begin{figure*}
	\begin{center}
		\includegraphics[scale=0.22]{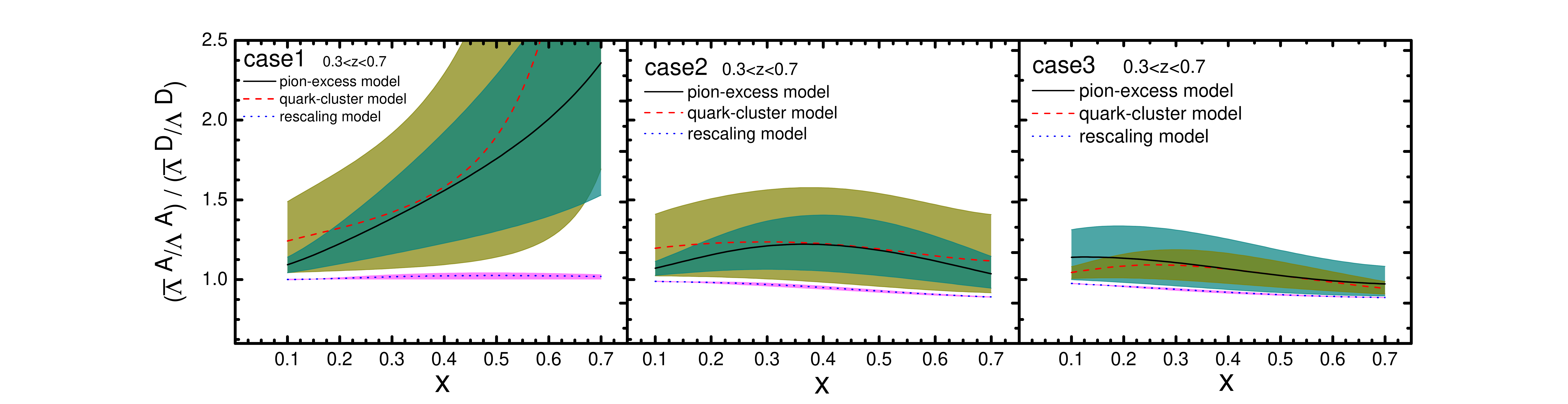}
	\end{center}

	\vspace{-0.5cm}
	\caption{\label{fig:25} The results of $(\bar{\Lambda}^{A}/\Lambda^{A})/(\bar{\Lambda}^{D}/\Lambda^{D})$ at $Q^{2}=5~\mathrm{GeV^{2}}$. The solid-black, dashed-red, and dotted-blue curves are the results of the pion excess model, the quark-cluster model, and the rescaling model. The nucleus $A$ is Fe. These three figures show different fragmentation function cases.}
\end{figure*}

\begin{figure*}
	\begin{center}
		\includegraphics[scale=0.22]{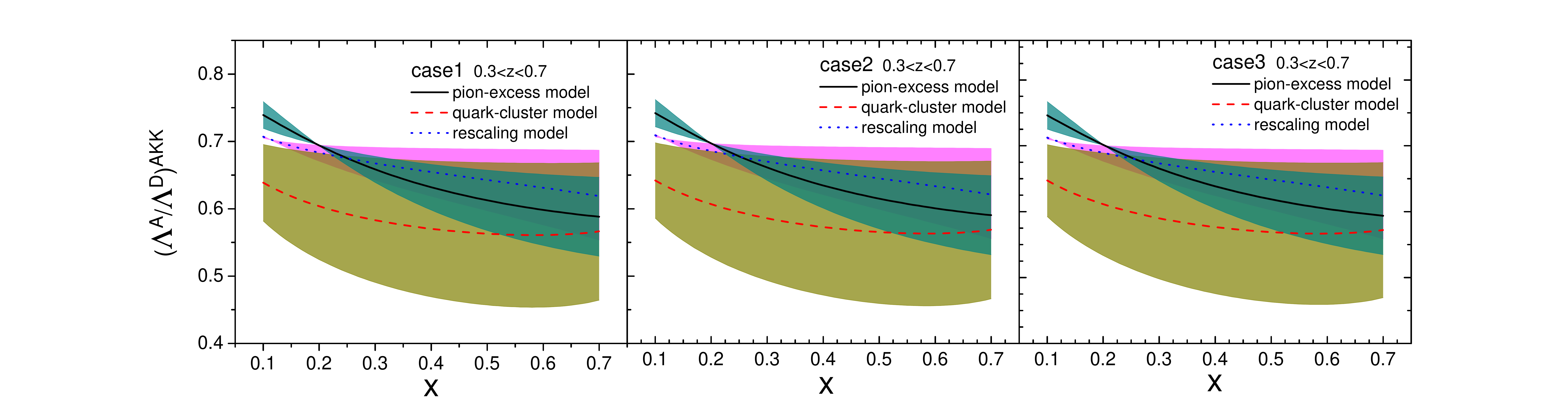}
	\end{center}
	\vspace{-0.5cm}
	\caption{\label{fig:26} The results of $(\Lambda^{A}/\Lambda^{D})^{\mathrm{AKK}}$ at $Q^{2}=5~\mathrm{GeV^{2}}$ considering the AKK~\cite{Albino:2008fy} modification. The solid-black, dashed-red, and dotted-blue curves are the results of the pion excess model, the quark-cluster model, and the rescaling model. The nucleus $A$ is Fe. These three figures show different fragmentation function cases.}
\end{figure*}

\begin{figure*}
	\begin{center}
		\includegraphics[scale=0.22]{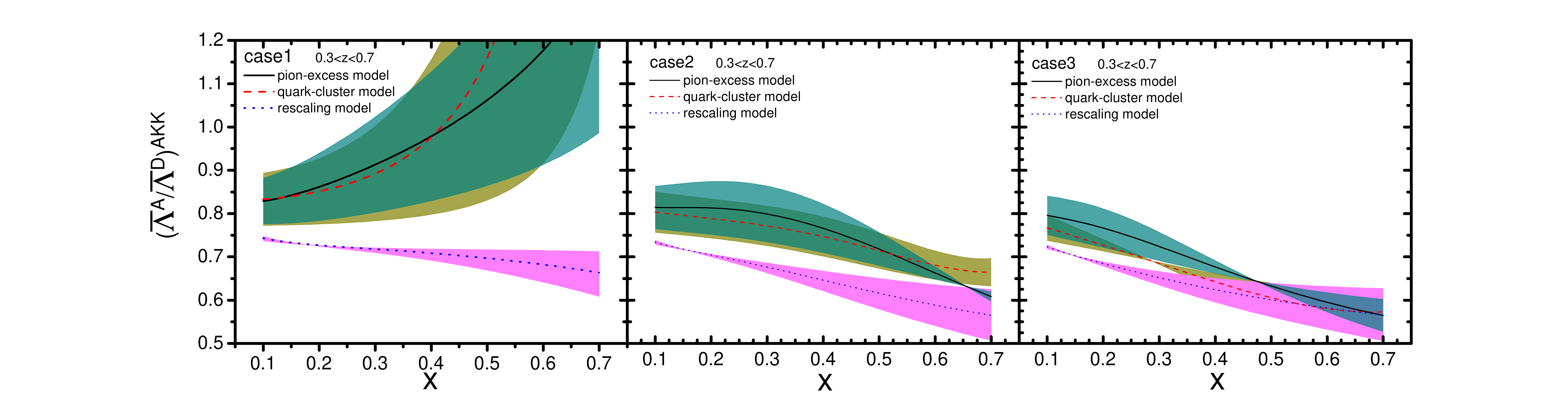}
	\end{center}
	\vspace{-0.5cm}
	\caption{\label{fig:27} The results of $(\bar{\Lambda}^{A}/\bar{\Lambda}^{D})^{\mathrm{AKK}}$ at $Q^{2}=5~\mathrm{GeV^{2}}$ considering the AKK~\cite{Albino:2008fy} modification. The solid-black, dashed-red, and dotted-blue curves are the results of the pion excess model, the quark-cluster model, and the rescaling model. The nucleus $A$ is Fe. These three figures show different fragmentation function cases.}
\end{figure*}

\begin{figure*}
	\begin{center}
		\includegraphics[scale=0.22]{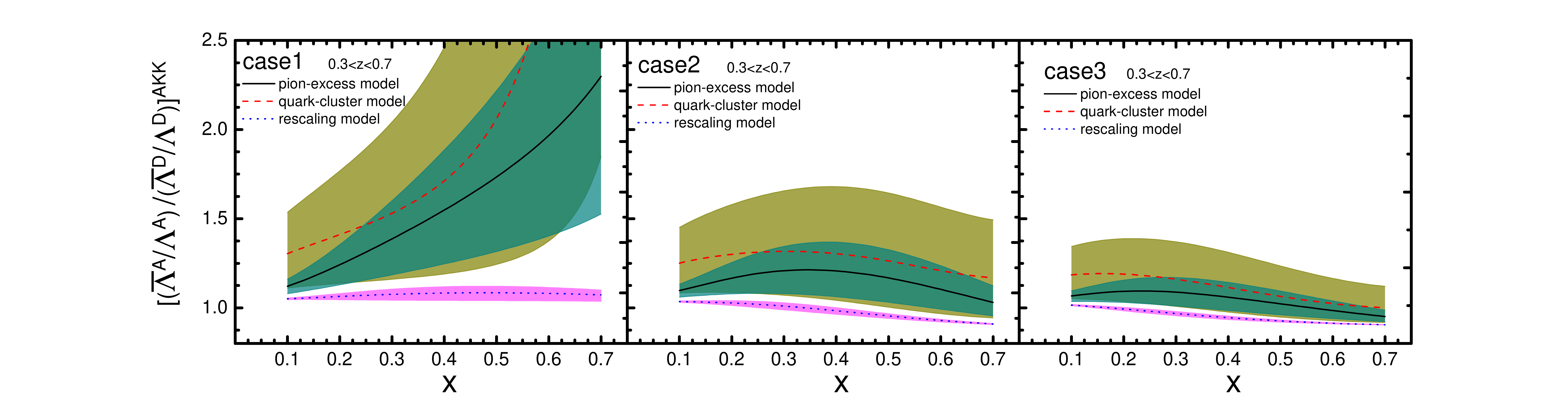}
	\end{center}
	\vspace{-0.5cm}
	\caption{\label{fig:28} The results of $\left[(\bar{\Lambda}^{A}/\Lambda^{A})/(\bar{\Lambda}^{D}/\Lambda^{D})\right]^{\mathrm{AKK}}$ at $Q^{2}=5~\mathrm{GeV^{2}}$ considering the AKK~\cite{Albino:2008fy} modification. The solid-black, dashed-red, and dotted-blue curves are the results of the pion excess model, the quark-cluster model, and the rescaling model. The nucleus $A$ is Fe. These three figures show different fragmentation function cases.}
\end{figure*}

In Figs.~\ref{fig:23},~\ref{fig:24}, and ~\ref{fig:25}, we show the behaviors of ratios $\Lambda^{A}/\Lambda^{D}$, $\bar{\Lambda}^{A}/\bar{\Lambda}^{D}$, and $(\bar{\Lambda}^{A}/\Lambda^{A})/(\bar{\Lambda}^{D}/\Lambda^{D})$. In Figs.~\ref{fig:26},~\ref{fig:27} and ~\ref{fig:28}, considering the AKK modification, we give the shapes of $(\Lambda^{A}/\Lambda^{D})^{\mathrm{AKK}}$, $(\bar{\Lambda}^{A}/\bar{\Lambda}^{D})^{\mathrm{AKK}}$, and $\left[(\bar{\Lambda}^{A}/\Lambda^{A})/(\bar{\Lambda}^{D}/\Lambda^{D})\right]^{\mathrm{AKK}}$. In each figure, we compare the distinct results by adopting different cases of fragmentation functions. The error bands in these figures reflect the uncertainties of the experimental data of the EMC ratio from the EMC collaboration~\cite{Aubert:1983xm} and from the BCDMS collaboration~\cite{Benvenuti:1987az}.

In Figs.~\ref{fig:23} and~\ref{fig:26}, we can see that the error bands overlap each other for the pion excess model, the quark-cluster model, and the rescaling model. The quantity $\Lambda^{A}/\Lambda^{D}$ has similar behaviors in these three different models. But in Figs.~\ref{fig:24} and ~\ref{fig:27}, we notice that different sets of quark fragmentation functions produce different shapes of $\bar{\Lambda}^{A}/\bar{\Lambda}^{D}$. This is because $\bar{\Lambda}$ is largely fragmented from the antiquarks inside the targets, so that the $x$ dependence of production ratio $\bar{\Lambda}^{A}/\bar{\Lambda}^{D}$ is sensitive to the sea quark distributions in the nucleus. However, in Figs.~\ref{fig:25} and ~\ref{fig:28} the ratio $(\bar{\Lambda}^{A}/\Lambda^{A})/(\bar{\Lambda}^{D}/\Lambda^{D})$ shows different shapes in three sets of fragmentation functions and three models. From these analyses, although the error bands of the quark-cluster model are wide, the $(\bar{\Lambda}^{A}/\Lambda^{A})/(\bar{\Lambda}^{D}/\Lambda^{D})$ ratio can discriminate the pion excess model, the six-quark cluster model, and the rescaling model to some extent. The quantity $(\bar{\Lambda}^{A}/\Lambda^{A})/(\bar{\Lambda}^{D}/\Lambda^{D})$ can serve as a tool to reveal sea quark components in the nucleus. These three models are not exactly resolved by available data, so we think that semi-inclusive scattering processes of $\Lambda$ and $\bar{\Lambda}$ productions are of significance for future research concerning the EMC effect.

\section{Summary}
In this paper, we make a brief review of three nuclear EMC models, namely the pion excess model, the quark-cluster model and the rescaling model. From the calculations on hadron production cross sections of $\Lambda$ and $\bar{\Lambda}$ both in the nucleus and in free nucleons, we show how different models of the EMC effect can be revealed by semi-inclusive deep inelastic scattering processes. The results show that the production of $\bar{\Lambda}$ is sensitive to the sea quark contributions in the nucleus. We therefore suggest that the quantity $(\bar{\Lambda}^{A}/\Lambda^{A})/(\bar{\Lambda}^{D}/\Lambda^{D})$ can provide some information to discriminate different models of the nuclear EMC effect.

We would like to thank Xiaozhen Du for helpful discussions. This work is partially supported by National Natural Science Foundation of China (Grant No.~11475006).



\begin{thebibliography}{56}
\bibitem{Aubert:1983xm}
  J.~J.~Aubert, {\it et al.} [European Muon Collaboration],
  Phys.\ Lett.\  {123B}, 275 (1983).


\bibitem{Arnold:1983mw}
  R.~G.~Arnold, {\it et al.},
  Phys.\ Rev.\ Lett.\  {52}, 727 (1984).


\bibitem{Aubert:1986yn}
  J.~J.~Aubert, {\it et al.} [European Muon Collaboration],
  Nucl.\ Phys.\ B {272}, 158 (1986).


\bibitem{Dasu:1988ru}
  S.~Dasu, {\it et al.},
  Phys.\ Rev.\ Lett.\  {60}, 2591 (1988).


\bibitem{Dasu:1993vk}
  S.~Dasu, {\it et al.},
  Phys.\ Rev.\ D {49}, 5641 (1994).


\bibitem{Gomez:1993ri}
  J.~Gomez, {\it et al.},
  Phys.\ Rev.\ D {49}, 4348 (1994).


\bibitem{Seely:2009gt}
  J.~Seely, {\it et al.},
  Phys.\ Rev.\ Lett.\  {103}, 202301 (2009)


\bibitem{Abramowicz:1984yk}
  H.~Abramowicz, {\it et al.},
  Z.\ Phys.\ C {25}, 29 (1984).


\bibitem{Parker:1983yi}
  M.~A.~Parker, {\it et al.} [BEBC TST Neutrino Collaboration],
  Nucl.\ Phys.\ B {232}, 1 (1984).


\bibitem{CooperSarkar:1984eb}
  A.~M.~Cooper-Sarkar, {\it et al.} (WA25 and WA59 Collaborations),
  Phys.\ Lett.\  {141B}, 133 (1984).


\bibitem{Ammosov:1984rd}
  V.~V.~Ammosov, {\it et al.},
  JETP Lett.\  {39}, 393 (1984)
  [Pisma Zh.\ Eksp.\ Teor.\ Fiz.\  {39}, 327 (1984)].


\bibitem{Hanlon:1985yg}
  J.~Hanlon, {\it et al.},
  Phys.\ Rev.\ D {32}, 2441 (1985).


\bibitem{Guy:1986us}
  J.~Guy, {\it et al.} (WA25 and WA59 Collaborations),
  Z.\ Phys.\ C {36}, 337 (1987).


\bibitem{Close:1984zn}
  F.~E.~Close, R.~L.~Jaffe, R.~G.~Roberts and G.~G.~Ross,
  Phys.\ Rev.\ D {31}, 1004 (1985).


\bibitem{McGaughey:1999mq}
  P.~L.~McGaughey, J.~M.~Moss and J.~C.~Peng,
  Ann.\ Rev.\ Nucl.\ Part.\ Sci.\  {49}, 217 (1999)


\bibitem{Chmaj:1983jq}
  T.~Chmaj and K.~J.~Heller,
  Acta Phys.\ Pol.\ B {15}, 473 (1984).


\bibitem{Chmaj:1985pp}
  T.~Chmaj and K.~J.~Heller,
  Lett.\ Nuovo Cimento\  {42}, 415 (1985).


\bibitem{DiasdeDeus:1984wy}
  J.~Dias de Deus, M.~Pimenta, and J.~Varela,
  Z.\ Phys.\ C {26}, 109 (1984).


\bibitem{DiasdeDeus:1984ge}
  J.~Dias de Deus, M.~Pimenta, and J.~Varela,
  Phys.\ Rev.\ D {30}, 697 (1984).


\bibitem{LlewellynSmith:1983vzz}
  C.~H.~Llewellyn Smith,
  Phys.\ Lett.\  {128B}, 107 (1983).


\bibitem{Ericson:1983um}
  M.~Ericson and A.~W.~Thomas,
  Phys.\ Lett.\  {128B}, 112 (1983).


\bibitem{Berger:1983jk}
  E.~L.~Berger, F.~Coester and R.~B.~Wiringa,
  Phys.\ Rev.\ D {29}, 398 (1984).


\bibitem{Pirner:1980eu}
  H.~J.~Pirner and J.~P.~Vary,
  Phys.\ Rev.\ Lett.\  {46}, 1376 (1981).


\bibitem{Jaffe:1982rr}
  R.~L.~Jaffe,
  Phys.\ Rev.\ Lett.\  {50}, 228 (1983).


\bibitem{Carlson:1983fs}
  C.~E.~Carlson and T.~J.~Havens,
  Phys.\ Rev.\ Lett.\  {51}, 261 (1983).


\bibitem{Close:1983tn}
  F.~E.~Close, R.~G.~Roberts and G.~G.~Ross,
  Phys.\ Lett.\  {129B}, 346 (1983).


\bibitem{Jaffe:1983zw}
  R.~L.~Jaffe, F.~E.~Close, R.~G.~Roberts and G.~G.~Ross,
  Phys.\ Lett.\  {134B}, 449 (1984).


\bibitem{Nachtmann:1983py}
  O.~Nachtmann and H.~J.~Pirner,
  Z.\ Phys.\ C {21}, 277 (1984).


\bibitem{Ma:2004zt}
  B.-Q.~Ma, I.~Schmidt and J.~J.~Yang,
  Phys.\ Lett.\ B {598}, 211 (2004)


\bibitem{Lu:2006xr}
  B.~Lu and B.-Q.~Ma,
  Phys.\ Rev.\ C {74}, 055202 (2006)


\bibitem{Chi:2013hka}
  Y.~Chi and B.-Q.~Ma,
  Phys.\ Lett.\ B {726}, 737 (2013)


\bibitem{Chi:2014xba}
  Y.~Chi, X.~Du and B.-Q.~Ma,
  Phys.\ Rev.\ D {90},  074003 (2014)


\bibitem{Du:2017nzy}
  X.~Du and B.-Q.~Ma,
  Phys.\ Rev.\ D {95}, 014029 (2017)


\bibitem{Benvenuti:1987az}
  A.~C.~Benvenuti, {\it et al.} [BCDMS Collaboration],
  Phys.\ Lett.\ B {189}, 483 (1987).


\bibitem{Dulat:2015mca}
  S.~Dulat, {\it et al.},
  Phys.\ Rev.\ D {93}, 033006 (2016)


\bibitem{Sutton:1991ay}
  P.~J.~Sutton, A.~D.~Martin, R.~G.~Roberts and W.~J.~Stirling,
  Phys.\ Rev.\ D {45}, 2349 (1992).


\bibitem{James:1975dr}
  F.~James and M.~Roos,
  Comput.\ Phys.\ Commun.\  {10}, 343 (1975).


\bibitem{Blankenbecler:1974tm}
  R.~Blankenbecler and S.~J.~Brodsky,
  Phys.\ Rev.\ D {10}, 2973 (1974).


\bibitem{Farrar:1975yb}
  G.~R.~Farrar and D.~R.~Jackson,
  Phys.\ Rev.\ Lett.\  { 35}, 1416 (1975).


\bibitem{Vainshtein:1977db}
  A.~I.~Vainshtein and V.~I.~Zakharov,
  Phys.\ Lett.\  {72B}, 368 (1978).


\bibitem{Sivers:1982wk}
  D.~W.~Sivers,
  Annu.\ Rev.\ Nucl.\ Part.\ Sci.\  {32}, 149 (1982).


\bibitem{deGroot:1978feq}
  J.~G.~H.~de Groot {\it et al.},
  Z.\ Phys.\ C {1}, 143 (1979).


\bibitem{Abramowicz:1982zr}
  H.~Abramowicz, {\it et al.},
  Z.\ Phys.\ C { 15}, 19 (1982).


\bibitem{Gribov:1971zn}
  V.~N.~Gribov and L.~N.~Lipatov,
  Phys.\ Lett.\  {37B}, 78 (1971).


\bibitem{Barone:2000tx}
  V.~Barone, A.~Drago and B.-Q.~Ma,
  Phys.\ Rev.\ C { 62}, 062201 (2000)


\bibitem{Ma:2003gd}
  B.-Q.~Ma, I.~Schmidt and J.~J.~Yang,
  Phys.\ Lett.\ B {574}, 35 (2003)

\bibitem{Ma:2001ri}
  B.-Q.~Ma, I.~Schmidt, J.~Soffer and J.~J.~Yang,
  Phys.\ Rev.\ D {65}, 034004 (2002)


\bibitem{Ma:1999gj}
  B.-Q.~Ma, I.~Schmidt and J.~J.~Yang,
  Phys.\ Lett.\ B {477}, 107 (2000)


\bibitem{Albino:2008fy}
  S.~Albino, B.~A.~Kniehl and G.~Kramer,
  Nucl.\ Phys.\ B {803}, 42 (2008)


\bibitem{Airapetian:2003mi}
  A.~Airapetian, {\it et al.} [HERMES Collaboration],
  Phys.\ Lett.\ B {577}, 37 (2003)


\bibitem{Bialas:1983kn}
  A.~Bialas and T.~Chmaj,
  Phys.\ Lett.\  {133B}, 241 (1983).


\bibitem{Bialas:1986cf}
  A.~Bialas and M.~Gyulassy,
  Nucl.\ Phys.\ B {291}, 793 (1987).


\bibitem{Accardi:2002tv}
  A.~Accardi, V.~Muccifora and H.~J.~Pirner,
  Nucl.\ Phys.\ A {720}, 131 (2003)


\bibitem{Wang:1996yh}
  X.~N.~Wang, Z.~Huang and I.~Sarcevic,
  Phys.\ Rev.\ Lett.\  {77}, 231 (1996)


\bibitem{Wang:2002ri}
  E.~Wang and X.~N.~Wang,
  Phys.\ Rev.\ Lett.\  {89}, 162301 (2002)


\bibitem{Arleo:2003jz}
  F.~Arleo,
  Eur.\ Phys.\ J.\ C {30}, 213 (2003)


\bibitem{Arleo:2002kh}
  F.~Arleo,
  JHEP {0211}, 044 (2002)




  \end{thebibliography}
\end{document}